\documentclass[sigconf]{acmart}

\AtBeginDocument{%
  }

\copyrightyear{2026}
\acmYear{2026}
\setcopyright{cc}
\setcctype{by-nc-nd}
\acmConference[CHI EA '26]{Extended Abstracts of the 2026 CHI Conference on Human Factors in Computing Systems}{April 13--17, 2026}{Barcelona, Spain}
\acmBooktitle{Extended Abstracts of the 2026 CHI Conference on Human Factors in Computing Systems (CHI EA '26), April 13--17, 2026, Barcelona, Spain}
\acmDOI{10.1145/3772363.3798837}
\acmISBN{979-8-4007-2281-3/2026/04}

\begin{document}


\title{From Uncertainty to Possibility: Early Computing Experiences for Rural Girls}

\author{Poornima Meegammana}
\email{hmee941@aucklanduni.ac.nz}
\orcid{0009-0005-1419-5747}
\affiliation{
  \institution{Faculty of Engineering and Design
University of Auckland}
  \city{Auckland}
  \country{New Zealand}
}

\author{Niranjan Meegammana}
\email{niranjan.meegammana@gmail.com }
\orcid{}
 \affiliation{
  \institution{Shilpa Sayura Foundation}
  \city{Kandy}
  \country{Sri Lanka}
}

\author{Chathurika Jayalath}
\email{madumali910@gmail.com}
\orcid{0009-0006-4573-8713}
 \affiliation{
  \institution{Foundation for Innovative Social Development}
  \city{Colombo}
  \country{Sri Lanka}
}
\author{Chethya Munasinghe}
\email{chethya.munasinghe@aod.lk}
\orcid{0009-0003-1545-8057}

 \affiliation{
  \institution{Academy of Design (AOD)}
  \city{Colombo}
  \country{Sri Lanka}
}

 \author{Kunal Gupta}
\email{kgup421@aucklanduni.ac.nz}
\orcid{}
 \affiliation{
  \institution{Empathic Computing Lab,
University of Auckland}
  \city{Auckland}
  \country{New Zealand}
}

\renewcommand{\shortauthors}{Trovato et al.}

\begin{abstract}
Girls remain underrepresented in computing, and rural contexts often compound barriers of access, language, and gender norms. Prior work in computing education highlights that confidence and belonging can shape participation, yet most evidence comes from well-resourced, English-dominant settings. Less is known about how locally grounded pathways can build programming self-efficacy and broaden career interest for adolescent girls. We addressed this gap by delivering a curriculum that began with digital foundations and unplugged problem-solving, then progressed to block-based programming activities, supported by parent awareness and teacher training in gender-responsive practices. Pre and post-surveys showed a reliable increase in programming self-efficacy, and career aspirations shifted toward technology. Complementary qualitative data indicate that mastery experiences, peer collaboration, and the creation of personal projects were key drivers of confidence, suggesting design priorities for scalable, locally relevant programmes in low-resource communities that can shift perceptions of who belongs in computing.
\end{abstract}


\begin{teaserfigure}
  \includegraphics[width=\textwidth]{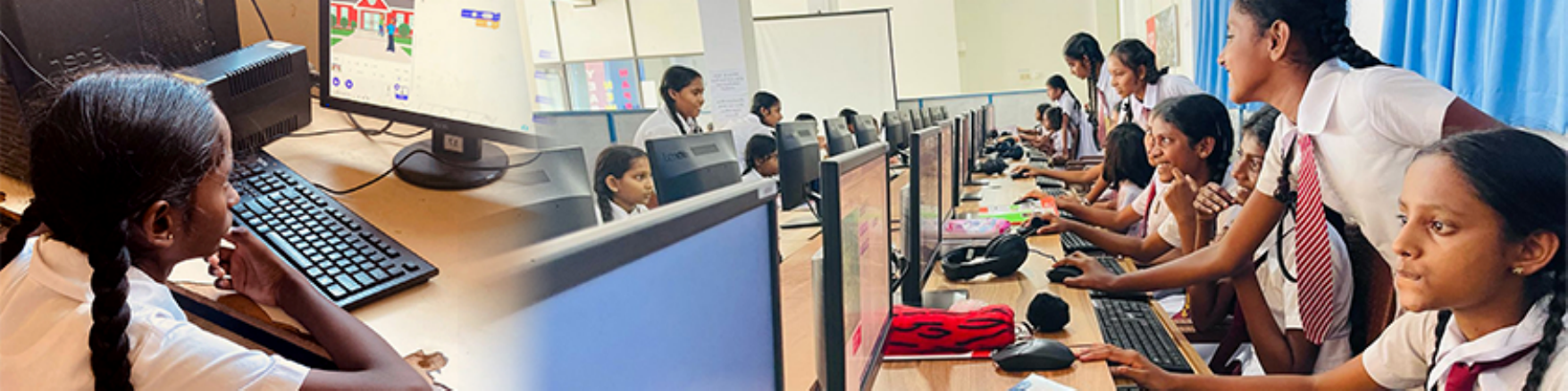}
  \caption{Girls working together during an introductory programming session. Permission to use these images has been granted.}
  \label{fig:teaser}
\end{teaserfigure}


\maketitle

\section{Introduction}
Women remain underrepresented in technology careers, with global participation below 20\% \cite{frachtenberg2022underrepresentation,main2017underrepresentation}. In Sri Lanka (SL), women comprise only 35\% of the ICT workforce \cite{staus2021addressing}. These disparities limit career opportunities, reduce income potential, and narrow the perspectives that shape technology \cite{campbell2025missing,convertino2020nuancing,centeno2025our,corbett2015solving}. Importantly, underrepresentation begins early. Gender gaps in self-efficacy (SE), interest, and exposure emerge during middle school and widen through adolescence, particularly in rural or under-resourced contexts where access is constrained, and gendered stereotypes around technology persist \cite{main2017underrepresentation,cheryan2017some}.

Rural girls often lack role models and early opportunities to engage with computing in culturally meaningful ways \cite{kelly2013willing,reed2022using}. Without early confidence-building experiences, many disengage before they can envision computing as a viable or relevant pathway. Prior research highlights the importance of self-efficacy and outcome expectations in shaping participation in computing, yet most evidence comes from well-resourced, English-dominant settings. Less is known about how to design confidence-building pathways for girls in low-resource, local-language contexts. In this work, we investigate how a locally grounded computing education programme shaped girls’ programming confidence and interest in technology careers in rural SL. We delivered an eight-week programme for adolescent girls that progressed from unplugged problem-solving to local-language, block-based programming. The curriculum scaffolded learners from abstract reasoning to creative programming through hands-on, personalised tasks, supported by girls-only learning environments, parent awareness sessions, and teacher training in gender-responsive practices. This work contributes empirical evidence from a low-resource, local-language context on how early computing experiences can support girls’ programming SE and emerging career interest. We translate these findings into design priorities for confidence-building computing education that foreground progression, creative ownership, language accessibility, and ecosystem support. Together, these insights extend existing discussions on inclusive computing education to rural and under-resourced settings. We ask:

\noindent
\textbf{RQ1.} How does participation in a computing education programme influence girls’ programming self-efficacy (SE) and outcome expectancy (OE) in low-resource contexts?

\noindent
\textbf{RQ2.} To what extent does participation in the programme influence girls’ interest in technology-related careers?

\section{Related work}
Social Cognitive Career Theory (SCCT) frames SE and outcome expectations as key mechanisms shaping students’ interests and career intentions \cite{lent1994toward}. In technology education, SE is a strong predictor of interest and persistence, while outcome expectations influence interest but are less consistently linked to behaviour \cite{turner2019ses,sheu2017scct}. Both constructs are shaped by early learning experiences and social context, including feedback, encouragement, and opportunities for early success \cite{nugent2015model,luo2021stem}. Learners who believe they can succeed in computing are more likely to anticipate positive outcomes from technology careers, such as fulfilment and social impact \cite{amalina2025factors,blotnicky2018study}. Early mastery experiences are particularly important for girls and first-time learners. Small early successes can produce lasting gains in programming SE, especially when supported by hands-on activities and peer interaction \cite{ramalingam2004self,wiedenbeck2004factors}. Block-based programming environments such as Scratch lower entry barriers by reducing syntactic complexity and supporting iterative exploration \cite{10.1145/1592761.1592779}. Unplugged computational thinking activities can further support abstraction and reasoning before learners transition to code, helping reduce anxiety in early stages of learning \cite{arslan2022investigation}. Language and cultural context also shape participation in computing. Learning to program in a second language can increase cognitive load and slow comprehension when instructional materials are primarily in English \cite{Guo2018Non-Native}. In contrast, instruction delivered in learners’ primary languages supports confidence and engagement \cite{dasgupta_hill_2017}. For rural girls, participation is further influenced by parental expectations, teacher beliefs, and local gender norms \cite{hand2017exploring,rozek2015gender}. Building on this, our study examines how a locally grounded computing programme, combining unplugged activities, local-language block-based programming, and supportive learning environments, shapes SE, OE, and technology-related career interest among rural girls in SL.

\section{Methodology}
We evaluated an eight-week computing programme delivered in collaboration with Shilpa Sayura Foundation (SSF) \cite{SSF_2009} in 
rural SL community training centres, using a within-subject pre to post design. Participants completed self-report measures of programming SE, OE, and career interest before and after the programme. To contextualise the questionnaire findings, we conducted semi-structured 30-minute interviews with 10 students and five 60-minute paired teacher interviews. Interviews were conducted in Sinhala and Tamil, audio-recorded, transcribed, and professionally translated into English, then analysed using reflexive thematic analysis [5]. \textbf{Participants:} We recruited 162 girls aged 10–14 years (M = 11.89, SD = 1.04) across five rural sites in SL: Agarapathana, Anuradhapura, Arachchikattuwa, Buttala and Tissamaharama, where access to formal computing education is limited. Most participants reported limited prior exposure to computing: 70.89\% did not have a computer at home, 52.53\% felt uncomfortable using a computer or tablet, and only 5.06\% reported feeling very comfortable. In addition, 63.29\% reported that they had never created a game or digital project using a computer or tablet.

\begin{figure}
    \centering
    \includegraphics[width=\linewidth]{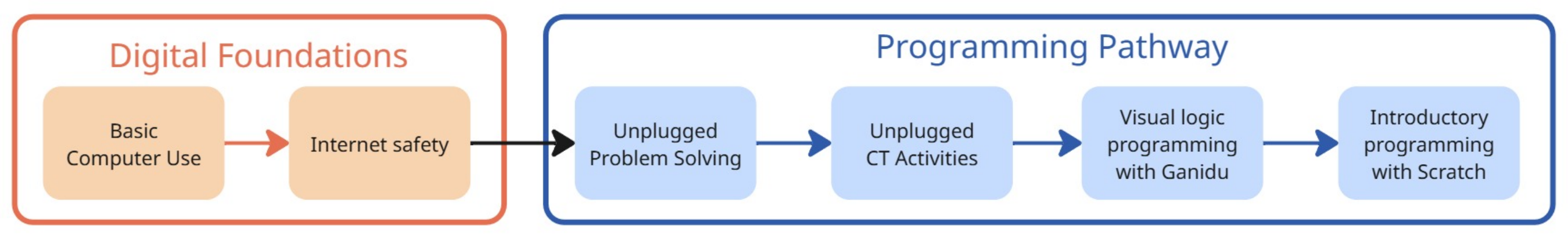}
    \caption{Curriculum pathway across Digital Foundations and the Programming Pathway}
    \label{fig:curriculum}
\end{figure}

\begin{figure}
    \centering
    \includegraphics[width=\linewidth]{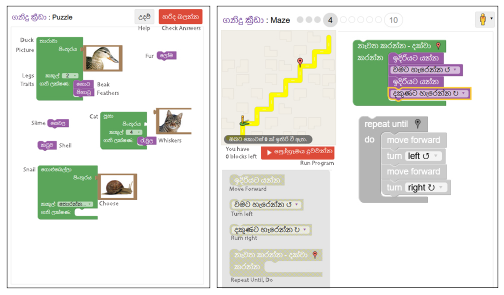}
    \caption{Ganidu interface showing local-language, block-based programming tasks.}
    \label{fig:Ganidu}
\end{figure}

\textbf{Curriculum and Learning Design:} The curriculum was organised into two phases. \emph{Digital Foundations} introduced basic computer use and online safety. The \emph{Programming Pathway} progressed from unplugged problem-solving and computational thinking activities to local-language block-based programming using Ganidu , followed by introductory programming in Scratch (Fig.~\ref{fig:curriculum}). This sequence supported a gradual transition from abstract reasoning to executable code. Ganidu is a local-language block-based programming tool developed by SSF, designed to introduce sequencing, repetition, and conditional logic through progressively challenging levels (Figure~\ref{fig:Ganidu})\cite{ganidu_2016}. Built on Google Blockly \cite{fraser_2014}, Ganidu served as a transition from unplugged activities to Scratch by enabling the translation of paper-based plans into executable steps. Scratch was used in later sessions as the primary programming environment due to its suitability for novice programmers \cite{10.1145/1592761.1592779,10.1145/1340961.1340974}. Core programming concepts included events, motion, loops, and conditionals.

This design was guided by contextual requirements identified through SSF's prior delivery at the same locations, where learners typically had minimal prior computing exposure and low English proficiency. We therefore used a progression from unplugged activities to Ganidu to Scratch, and delivered instruction primarily in students’ first language while selectively introducing essential computing vocabulary encountered in the tools.

\textbf{Study Procedure:} Ethical approval was obtained from the Research Ethics Committee for Social Sciences and Humanities, University of Colombo.
Local training centres promoted the programme. Interested students registered through the centres, then each site held parent meetings with sessions on technology careers and how family support enables girls’ participation in computing. Each centre had two facilitators trained in the curriculum and gender-responsive teaching by a gender expert.
The programme ran for eight weeks, with one two-hour session per week. All instructions, interviews, and questionnaires were administered in Sinhala and Tamil. We enacted gender-responsive teaching through multiple facilitation routines, including girls-only sessions led by female facilitators. Examples, scenarios, and characters in demonstrations and practice tasks centred girls and women to counter computing stereotypes. Feedback emphasised effort, strategy, and progress rather than speed or correctness. Pre- and post-programme questionnaires were collected to measure programming SE, OE, and career interest. Demographic information was collected at baseline. SE and OE were measured using items adapted from the Computer Science Attitudes Scale \cite{RACHMATULLAH2020100018}. Career interest was captured through an open-ended question and coded as technology-related or non-technology-related.

\textbf{Data Analysis:} Data were collected from 162 participants. Four responses were excluded due to missing data, resulting in a final sample of 158 participants. Normality of SE and OE scores was assessed using the Shapiro–Wilk test. As the distributions were non-normal, Wilcoxon signed-rank tests were used to examine pre–post changes in these measures. Career interest responses were analysed as a binary outcome (technology-related vs. non-technology-related). A McNemar test was used to examine changes in the proportion of participants reporting technology-related career aspirations from pre to post.

\section{Results}
\begin{figure}
    \centering
    \includegraphics[width=\linewidth]{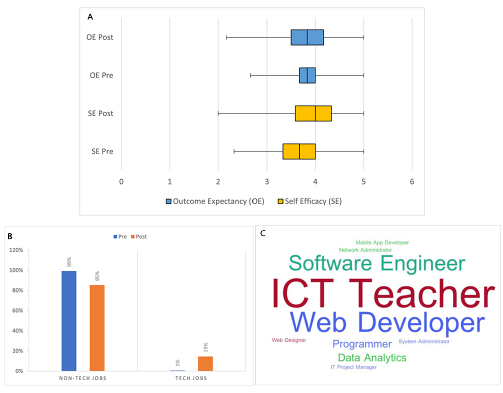}
    \caption{(A) Self-efficacy and outcome expectancy scores (boxplots), (B) Career aspirations by type (pre/post), (C) Post-programme technology career mentions (word cloud)}
    \label{fig:results}
\end{figure}

We examined changes in self-efficacy, Outcome expectancy, and career interest. As the measures were non-normally distributed, Wilcoxon signed-rank tests were used for pre–post comparisons. \textbf{Programming self-efficacy:} Increased significantly from pre to post ($Z = -2.913, p = 0.003$). Mean increased from pre (M = 3.72, SD = 0.49) to post (M = 3.89, SD = 0.62) (Figure~\ref{fig:results}A). \textbf{Outcome expectancy:} Did not change significantly over the programme ($Z = 0.295$, $p = 0.77$), with similar mean scores at pre (M = 3.89, SD = 0.46) and post (M = 3.86, SD = 0.51). \textbf{Career interest:} This was analysed as a binary outcome (technology-related vs. non-technology or undecided). A significantly greater proportion of participants reported technology-related career aspirations at post (McNemar test, $\chi^2 = 5.907$, $p = 0.015$). Several participants shifted from non-technology to technology-related careers, with no shifts in the opposite direction (Figure~\ref{fig:results}B). Before the programme, top career aspirations were undecided (30.4\%), doctor (29.7\%) and teacher (29.1\%) with only one mention of a programming-related career. In post, undecided responses dropped to 22.8\%, and aspirations became more specific, with emerging interest in technology roles such as ICT teacher (3.2\%), software engineer (2.5\%), and web developer (2.5\%)
(. Figure~\ref{fig:results}C illustrates the range of technology roles reported.

\section{Discussion}
We observed a statistically significant increase in girls’ programming SE from pre to post. Although the magnitude of change was modest, it is meaningful given participants’ limited prior exposure to computing and constrained access. Students described growing confidence through completing projects, debugging, and extending their work beyond initial tasks. One student reflected, \textit{``After I made the simple version, I tried adding new things, following a YouTube video.''} Teachers similarly noted a progression from reliance on step-by-step guidance to more independent experimentation. These accounts align with the idea that early mastery experiences support confidence, consistent with prior work on SE in computing education~\cite{banduraself-efficacy1997}. Social interaction further shaped these experiences. Students frequently worked through challenges together, and teachers described a norm of peer assistance, noting that \textit{``students who do well help other students.''} Such collaborative problem-solving and exposure to nearby peers’ success can support persistence and self-belief, particularly for novice learners \cite{Liu2023Understanding, Kleppang2023Explaining, Shao2022The}. Extending prior findings \cite{tellhed2022sure, smit2025experiencing}, our results 
show similar patterns in a rural, low-resource context, consistent with the idea that structured, hands-on learning may support confidence.

Outcome expectancy did not change significantly from pre to post. This is plausible given that many participants entered with strong beliefs about the value of technology (e.g., \textit{``Technology will be needed for my future job''}), leaving limited room for short-term change, a pattern also reported in after-school STEM programmes where participation may already reflect positive utility beliefs \cite{staus2021addressing,10.1145/3724363.3729037}. Teachers also highlighted structural constraints, including limited device access and connectivity outside programme sessions, which may restrict how learners imagine longer-term computing opportunities \cite{warschauer2004technology}.

However, we also observed an emerging shift in career interests, with 15\% of participants moving from undecided or non-tech aspirations toward technology-related goals. Notably, this shift was characterised less by the replacement of existing aspirations and more by a reduction in uncertainty: careers such as doctor and teacher remained common, while undecided responses decreased and specific technology-related roles (e.g., ICT teacher, software engineer, web developer) emerged post-programme. Students described reduced fear and increased independence (\textit{“My fear of computers has disappeared”}), while teachers noted greater willingness to experiment and persist (\textit{“Now girls are doing things on their own”}). Students and teachers linked enjoyment and pride in personalised projects with continued engagement. 
This pattern aligns with SCCT \cite{lent1994toward}, in which self-efficacy supports the development of interests.

\subsection{Design Implications}
Our findings suggest design strategies for inclusive, confidence-building computing education in low-resource settings. \newline
\textbf{Progressive scaffolding across representations}:
Teachers described the unplugged activities to Ganidu and then  Scratch, sequence as a helpful progression for learners with no prior experience. They noted that Ganidu eased the transition to Scratch by reducing cognitive load and helping students map abstract plans to executable steps. Designing for continuity across representations can support confidence as learners move toward more complex programming environments. \newline \textbf{Creative ownership through personal expression}: Open-ended, personalised projects were associated with high engagement in student and teacher accounts. Students described adding preferred shapes, sounds, and stories, while teachers highlighted strong interest in storytelling and animation. 
Participants described self-expression as a reason they connected tasks to their interests. \newline \textbf{Language-first access to computation}: 
Students and teachers reported that local-language, lowered perceived barriers for understanding in contexts where English proficiency is low. Students frequently relied on translations for programming terms (e.g., \textit{forever}), and some reported increased confidence in English through repeated exposure in meaningful tasks. Designing programming environments that foreground primary-language access can support comprehension without isolating learners from dominant technical vocabularies. \newline \textbf{Socially supported independence}:
Teachers described a learning culture that normalised collaboration and help-seeking, which may have contributed to persistence. Students often worked through challenges together before approaching facilitators, reducing fear of failure and reinforcing shared problem-solving. Across the programme, teachers reported that repeated hands-on practice supported a shift toward greater independence.
Designing for peer-supported progression can help learners move from dependence to self-directed exploration at home. \newline \textbf{Making technology futures legible}: For young learners with limited exposure to computing, designs should support exploration of technology-related activities without requiring alignment to a single career outcome. Providing examples that span creative, educational, and applied uses of computing (e.g., storytelling, teaching, problem-solving) can help learners map computing to familiar roles and interests, allowing career ideas to emerge gradually rather than forcing early decisions. \newline \textbf{Ecosystem alignment beyond the classroom}: 
Parents reported greater recognition of computing’s value, while teachers described increased student participation and a greater willingness to experiment. Although not tested separately, delivering the programme in girls-only groups may have further supported comfort and participation, with teachers noting greater confidence to speak up without fear of judgement. Designing programmes that align classroom experiences with family and school expectations can strengthen belonging and sustain confidence over time. Taken together, these implications emphasise design features that may help support confidence and aspirations in technology.

\section{ Limitations and Future Work}
This study had several limitations. Without a control group, observed changes cannot be attributed solely to the programme. Outcomes were self-reported and may reflect social desirability, differences in interpretation, and novelty or facilitator effects in a setting with limited prior exposure to computing. Time constraints may have limited deeper practice in debugging, and some programming terms remained in English, which may have affected students' comprehension of the task for those with lower English proficiency. Future work will examine whether gains are sustained through longitudinal delivery, add performance-based measures of computational thinking and programming understanding, and strengthen links to culturally relevant technology pathways through local role models and real-world applications.


\begin{acks}

The study was funded by the Fund for Innovation and Transformation, a program of the Inter-Council Network administered by the Manitoba Council on International Cooperation and using funds from Global Affairs Canada.

\end{acks}


\bibliographystyle{ACM-Reference-Format}
\bibliography{sample-base}

\appendix

\end{document}